\documentclass[preprint,12pt,nopreprintline]{elsarticle}

\usepackage{amssymb}

\usepackage{amsmath}

\usepackage{booktabs}
\usepackage{graphicx}
\usepackage{epsfig}
\usepackage{dcolumn}
\usepackage{multirow}
\usepackage{slashed}
\usepackage[colorlinks,linkcolor=red,anchorcolor=green,citecolor=blue]{hyperref}

\journal{Physics Letters B}
\graphicspath{{figs/}}
\hypersetup{pdftitle={Scalar glueball-s-sbar mixing in one flavor lattice QCD},
 pdfauthor={Long-Cheng Gui, Wei Sun, Ying Chen, Ming Gong, Geng Li, Zhaofeng Liu}}

\begin{document}

\begin{frontmatter}

 \title{Scalar glueball-$s\bar{s}$ mixing in one flavor lattice QCD}

 \author[lab1,lab2]{Long-Cheng Gui \corref{cor1}}
\ead{guilongcheng@hunnu.edu.cn}

 \author[lab3]{Wei Sun}

\author[lab3,lab4]{Ying Chen   \corref{cor1} }
\ead{cheny@ihep.ac.cn}

\author[lab3,lab4]{Ming Gong}

\author[lab3,lab4]{Geng Li}

\author[lab3]{Zhaofeng Liu}

\cortext[cor1]{Corresponding authors.}

\affiliation[lab1]{Department of Physics, Hunan Normal University, and Key Laboratory of Low-Dimensional Quantum Structures and Quantum Control of Ministry of Education, Changsha 410081, China}
\affiliation[lab2]{Synergetic Innovation Center for Quantum Effects and Applications (SICQEA), Hunan Normal University, Changsha 410081, China}
 \affiliation[lab3]{Institute of High Energy Physics, Chinese Academy of Sciences, Beijing 100049, China}
\affiliation[lab4]{School of Physical Sciences, University of Chinese Academy of Sciences, Beijing 100049, China}

\begin{abstract}

We investigate the mixing between the lowest-lying scalar glueball and the $s\overline{s}$ meson in $N_f=1$ lattice quantum chromodynamics (QCD) utilizing an anisotropic $16^3 \times 128$ lattice ensemble at a lattice spacing $a_s\approx 0.148\,\rm{fm}$. By solving a generalized eigenvalue problem (GEVP) for the optimized glueball and $s\overline{s}$ scalar operators in the $J^\text{PC} = 0^{++}$ channel, the masses of the two lowest-lying eigenstates are determined to be $m_1 = 1.296(20)\,\mathrm{GeV}$ and $m_2 = 1.768(48)\,\mathrm{GeV}$. By extracting the couplings of these mass eigenstates to the glueball and $s\bar{s}$ operators, we determine a substantial mixing angle $|\theta| = 40.5(2.2)^\circ$. Within a phenomenological two-state mixing model, we obtain a model-dependent mixing energy $|x_s| = 231(14)$ MeV. The quoted uncertainties combine statistical errors and those from the analysis procedure variation. These results indicate a strong glueball-$s\bar{s}$ mixing in the scalar sector, providing important non-perturbative inputs for understanding the nature of the experimental isoscalar scalar mesons. The continuum limit of the mixing energy and its quark mass dependence need to be investigated in the future.
\end{abstract}

 \end{frontmatter}

 \section{Introduction}
The quark model describes mesons as quark-antiquark ($q\bar{q}$) bound states. Since gluons are also the fundamental degrees of freedom of QCD, it is usually conjectured that there also exist bound states made up of gluons, namely, glueballs.  The scalar glueball is particularly important in QCD because the scalar gluonic operator is directly related to the trace anomaly of the energy-momentum tensor. As a predominantly gluonic massive excitation, it therefore provides a unique window into the emergence of the nonperturbative mass scale in QCD~\mbox{
\cite{Suganuma2023}}\hskip0pt
.  Lattice QCD studies show the existence of pure gauge glueballs in the quenched approximation and predict their spectrum~\cite{Morningstar:1997ff, Morningstar:1999rf, Chen:2005mg}.
If the $q\bar{q}$ mesons and pure gauge glueballs are taken as conceptual bare states, the quark-gluon dynamics in QCD will result in the mixing between $q\bar{q}$ mesons and glueballs with the same quantum numbers. In this sense, physical meson states can be mass eigenstates after mixing, and the mixing strength is crucial for determining the meson-state wave functions when glueball-$q\bar{q}$ mixing is considered.

The $q\bar{q}$ meson-glueball mixing in the scalar channel is intriguing. Lattice QCD studies predict the mass of the lowest-lying scalar glueball to be around $1.5-1.8\,\mathrm{GeV}$~\cite{Morningstar:1997ff, Morningstar:1999rf, Chen:2005mg, Richards:2010ck, Gregory:2012hu, Sun:2017ipk, Chen:2021dvn}. Experimentally, there are three isoscalar scalar mesons of masses around $1.5\,\mathrm{GeV}$, namely, $f_0(1370)$, $f_0(1500)$ and $f_0(1710)$~\cite{ParticleDataGroup:2024cfk}. These states can be sorted into a flavor SU(3) $q\bar{q}$ nonet of spin parity quantum numbers $J^P=0^+$ along with $K_0^*(1430)$ and $a_0(1450)$. However, one surplus isoscalar state here hints at the existence of an additional degree of freedom, namely, a scalar glueball state, since the quark model permits only two $q\bar{q}$ isoscalars in this mass region. In other words, the three $f_0$ states can be admixtures of the glueball state and the two $q\bar{q}$ isoscalars. Numerous phenomenological studies have investigated the mixing between the glueball and scalar $q\bar{q}$ mesons~\cite{Cheng:2006hu,Cheng:2008ss,Cheng:2009dg,Cheng:2015iaa,Vento:2015yja,Zhang:2016vcx,Guo:2020akt} based on the various theoretical assumptions; however, no consensus has been achieved yet on the mixing pattern.

The first lattice QCD study on the scalar glueball-$q\bar{q}$ mixing appeared two decades ago in the quenched approximation~\cite{Lee:1999kv}. The mixing energy of the scalar glueball-$s\bar{s}$ meson is measured to be around 40 MeV in the continuum limit. A follow-up quenched lattice QCD study on a coarse lattice obtains a much larger mixing energy of roughly 200-300 MeV after considering the quark annihilation diagram~\cite{McNeile:2000xx}. The quenched lattice QCD studies do not consider the sea quark effects, namely, the contribution from sea quark loops, such that the propagator of a flavor singlet meson is not complete even though the annihilation diagrams of valence quarks are included. The absence of sea quark loops may seriously distort the theoretical information conveyed by the propagator, especially in the flavor singlet scalar channel (and the pseudoscalar channel). So it is desired to study the properties of scalar mesons through lattice QCD in the presence of light dynamical quarks. However, the presence of light sea quarks makes the situation very complicated since the spectrum in the scalar channel is composed of many scalar mesons below $2.0\,\mathrm{GeV}$ as well as many multi-hadron states. Recent lattice QCD calculations have started to investigate glueball-meson mixing in the presence of dynamical quarks. In particular, the scalar glueball-charmonium mixing has been studied in $N_f=2$ QCD using an enlarged operator basis, where the overlaps between the physical states and the gluonic/quark bilinear operators were analyzed to quantify the mixing structure~\cite{Urrea-Nino:2026cjj}. Nevertheless, a systematic and comprehensive full-QCD lattice study on scalar mesons near the physical point ($N_f=2+1$ QCD at the physical pion and kaon masses) is still out of reach in the present era~\cite{Brett:2019tzr}.

To focus on the nonperturbative $q\bar{q}$-gluon dynamics, we can simplify the question by performing a lattice QCD study with one flavor of the dynamical quark, whose mass parameter can be tuned to make the lowest scalar mesons stable. In this case, the spectrum of scalar mesons is sparse in the low-energy region and facilitates us in studying the related glueball-$q\bar{q}$ mixing problem. The QCD with one flavor of dynamical strange quark ($s$) seems to meet this requirement. Previous lattice QCD studies show that the pseudoscalar $s\bar{s}$ meson has a mass around 690 MeV~\cite{Davies:2009tsa,Woss:2020ayi,Shi:2024fyv,CLQCD:2023sdb,Borsanyi:2020mff,Boccaletti:2024guq} without considering the annihilation diagram. When the annihilation diagram contribution is included for one flavor  of  strange quark, the pseudoscalar $s\bar{s}$ meson ($\eta_{(1)}$) mass is around 783 MeV~\cite{Shi:2024fyv}. So in this work, we use the lattice gauge ensemble with one flavor of strange quark~\cite{Shi:2024fyv} to explore the mixing between the lowest-lying scalar glueball and the $s\bar{s}$ meson. The method is similar to that used in the study of $\eta-\eta'$ mixing~\cite{Christ:2010dd,Dudek:2013yja}, the $\eta_c$-glueball mixing~\cite{Zhang:2021xvl}, and $\eta$-glueball mixing~\cite{Jiang:2022ffl}. The distillation method~\cite{Peardon:2009gh} is adopted for the treatment of the gauge covariant smearing of quark fields and the all-to-all quark propagators (perambulators) (and therefore the quark annihilation effects).

This paper is organized as follows: Sec.~\ref{sec:ensemble} presents the details of the $N_f=1$ gauge ensemble. In Section~\ref{sec:mixing1}, we present the numerical details in deriving the glueball-$s\bar{s}$ meson mixing angle. The phenomenological implication of the results will be discussed in Sec.~\ref{sec:discussion}. Sec.~\ref{sec:summary} is a brief summary.

\section{Numerical details}\label{sec:ensemble}

We generate gauge configurations with $N_f=1$ dynamical strange quarks on an $L^3\times T=16^3\times 128$ anisotropic lattice. We use the tadpole-improved Symanzik's gauge action for anisotropic lattices~\cite{Morningstar:1997ff,Chen:2005mg} and the tadpole-improved anisotropic clover fermion action~\cite{Edwards:2008ja,Sun:2017ipk}. The RHMC algorithm implemented in Chroma software~\cite{Edwards:2004sx} is used to generate the $N_f=1$ gauge configurations. The parameters in the action are tuned to give the anisotropy $\xi=a_s/a_t\approx 5$, where $a_t$ and $a_s$ are the temporal and spatial lattice spacings, respectively.   The  bare strange quark mass parameter   $a_t m_s^{\rm bare}$ is tuned using  the ratio $m_\phi/m_{\eta_s}=1.487$, where $m_\phi\approx 1020~\mathrm{MeV}$ is the experimental mass of the vector meson $\phi$ and $m_{\eta_{s}}=686(4)~\mathrm{MeV}$ is the mass value of the $s\bar{s}$ pseudoscalar meson (from only the connected quark diagram) that is determined by the HPQCD collaboration~\cite{Davies:2009tsa}. Then we use $m_\phi^2-m_{\eta_s}^2\approx 0.57(1)~\mathrm{GeV}^2$ to set the scale parameter $a_t^{-1}\approx 6.66(5)~\mathrm{GeV}$~\cite{Jiang:2022ffl,Shi:2024fyv}. Finally, we obtain $m_{\eta_s}=693.1(3)(6.0)\,\mathrm{MeV}$ and $m_\phi=1027.2(5)(7.7)\,\mathrm{MeV}$ on our gauge ensemble, where the second errors come from the uncertainty of $a_t$. This serves as a self-consistent check of our lattice setup. The details of the gauge ensemble are given in Table~\ref{tab:config}.
 \begin{table}[t]
    \centering
    \renewcommand\arraystretch{1.5}
    \caption{Parameters of the gauge ensemble.}
    \label{tab:config}
        \begin{tabular}{lcccccc}
            \toprule
            $L^3 \times T$    & $\beta$ & $a_t^{-1}\,\mathrm{(GeV)}$ & $\xi$      & $m_{\eta_s}\,\mathrm{(MeV)}$ & $m_\phi\,\mathrm{(MeV)}$  & $N_\mathrm{cfg}$ \\
            \midrule
            $16^3 \times 128$ & $2.4$     & $6.66(5)$       & $\sim 5.0$ & $693(6)$          & $1027(8)$      & $4893$           \\
            \bottomrule
        \end{tabular}
\end{table}

In $N_f=1$ QCD, all the $q\bar{q}$ mesons are flavor singlets, whose correlation functions receive  contributions from both the connected and disconnected quark diagrams. In the practical calculation, we adopt the distillation method \cite{Peardon:2009gh}, which facilitates a systematic treatment of the all-to-all quark propagators and smeared quark interpolation operators. On each timeslice of each configuration, we calculate $N_V=70$ eigenvectors $\{V_i(\vec{x},t),i=1,2,\ldots, N_V\}$ of the gauge covariant Laplacian operator, which span a Laplacian Heaviside subspace (LHS). The perambulators of the strange quark, which encapsulate the all-to-all quark propagators, are calculated in the LHS. The $N_V$ eigenvectors also provide a LHS smearing scheme for the quark field, namely, $\psi^{(s)}(\vec{x},t)=\sum_i V_i(\vec{x},t)V_i^*(\vec{y},t)\psi(\vec{y},t)$, where $\psi^{(s)}$ is the LHS smeared quark field. Throughout this work, meson operators are built in terms of $s^{(s)}$ and $\bar{s}^{(s)}$ fields, and the superscripts are omitted for convenience in the following discussions.

\begin{table}[t]
    \centering
    \renewcommand\arraystretch{1.5}
    \caption{The mass spectrum of all the mesons whose masses are below the expected decay thresholds. The mass $m^{(C)}$ refers to that from the connected diagram only and $m$ is derived after considering both the connected and disconnected diagrams. The mass values are converted by $a_t^{-1}=6.66\,\rm{GeV}$ and the errors are only statistical without including the uncertainty of $a_t^{-1}$.}
    \label{tab:spectrum}
        \begin{tabular}{ccrcc}
            \toprule
            name         & $J^\text{PC}$   & Decay mode                      &  $m^{(C)}$ (MeV)  & $m$ (MeV)  \\
            \midrule
            $\eta_{(1)}$ &  $0^{-+}$  &                               &  693.1(3)         &  783.0(5.5)    \\
            $ \phi$      &  $1^{--}$  &   $\eta_{(1)}\eta_{(1)}(L=1)$   &  1027.2(5)          &  1026(10)      \\
            $ f_0$       &  $0^{++}$  &   $\eta_{(1)}\eta_{(1)}(L=0)$   &  1486(2)          &  1287(11)      \\
            $ f_1$       &  $1^{++}$  &   $\phi\phi(L=2)$               &  1564(2)          &  1572(22)      \\
            $ h_1$       &  $1^{+-}$  &   $\eta_{(1)}\phi(L=0)$         &  1589(2)          &  1577(40)      \\
            $ f_2$       &  $2^{++}$  &   $\eta_{(1)}\eta_{(1)}(L=2)$   &  1629(3)          &  1642(22)      \\
            $ \phi_2$    &  $2^{--}$  &   $\eta_{(1)}\phi(L=1)$         &  2030(9)          &  2035(11)      \\
            $\eta_2$     &  $2^{-+}$  &   $\phi\phi(L=1)$               &  2045(11)         &  2044(19)      \\
            \bottomrule
        \end{tabular}
\end{table}

The lightest meson in our lattice setup is the pseudoscalar meson $\eta_{(1)}$ (we use this name to differentiate from $\eta_s$ that is used to set the scale) with a mass $m_{\eta_{(1)}}=783.0(5.5)~\mathrm{MeV}$~\cite{Shi:2024fyv} after the disconnected diagram is considered.
As a calibration, we calculate the mass spectrum of all the mesons whose masses are below the expected decay thresholds. In order to check the disconnected diagram contribution to the propagation of different mesons, for each meson, we extract the mass value $m^{(C)}$ from the connected diagram and the full-correlator mass $m$ by also including the disconnected diagram. The results are shown in Table~\ref{tab:spectrum}, where
the values in physical units are converted using $a_t^{-1}=6.66~\rm{GeV}$ and errors are statistical without considering the uncertainty of $a_t$. The third column of Table~\ref{tab:spectrum} also shows the two-body decay modes of the lowest mass thresholds for specific $J^\text{PC}$ quantum numbers. With the obtained masses $m$ of $\eta_{(1)}$ and $\phi$, it is seen that most of the mesons in the table lie below the corresponding two-body thresholds except for $f_2$ and $\phi_2$ and are therefore stable particles under our lattice setup. Since the decays $f_2\to\eta_{(1)}\eta_{(1)}$ and $\phi_2\to \eta_{(1)}\phi$ can take place only in the $D$-wave and $P$-wave, respectively, their decay widths are expected to be small owing to the centrifugal barriers and the small decay phase spaces. So they can also be taken as quasi-stable particles.

Benefiting from the large statistics, the statistical errors of $m^{(C)}$ for these mesons are determined to be in the sub-percent level. In the quark model picture, the $h_1$ and $f_{0,1,2}$ states can be sorted into the $1P$ $s\bar{s}$ supermultiplet, namely, the $n^{2S+1}L_J=(1^1P_1, 1^3P_{0,1,2})$ multiplet. If $m^{(C)}$ of these states are treated as the masses of the pure $s\bar{s}$ mesons, the mass splitting pattern of this multiplet is very similar to that of its charmonium counterpart, namely, the $(h_c,\chi_{c0,1,2})$ multiplet. It is interesting to see that, if we introduce the $1P$ center-of-gravity (COG) mass similar to the charmonium and bottomonium cases
\begin{equation}
    m_\mathrm{COG}^{(C)}=\frac{1}{9}\left(m_{f_0}^{(C)}+3m_{f_1}^{(C)}+5m_{f_2}^{(C)}\right),
\end{equation}
then we have $m_\mathrm{COG}^{(C)}=1591(2)\,\rm{MeV}$, which is almost the same as $m_{h_1}^{(C)}=1589(2)\,\rm{MeV}$. Although we are not sure whether this is accidental or not, this relation is exactly the quark model expectation for $1P$ $q\bar{q}$ bound states and is supported by the experimental spectra of $1P$ charmonia and $1P$ bottomonia to a very high precision~\cite{ParticleDataGroup:2024cfk}.

Taking a look at the full-correlator masses $m$ in Table~\ref{tab:spectrum}, the inclusion of disconnected diagrams changes the mass values of $\eta_{(1)}$ and $f_0$ significantly, but almost does not affect the mass values of other states except for enlarging the statistical errors. The $\eta_{(1)}$ mass change can be understood by the Witten-Veneziano mechanism~\cite{Witten:1978bc,Witten:1979vv,Veneziano:1979ec} that the QCD $\mathrm{U}_A(1)$ anomaly introduces nonperturbative gluonic interactions between gluons and flavor-singlet pseudoscalars, which lifts the mass of $\eta_{(1)}$ a lot above $m_{\eta_s}$. It is intriguing to explore the dynamics behind the large $f_0$ mass change ($\sim 200$ MeV) when the $s\bar{s}$ annihilation effect is taken into account.

In the presence of $N_f=1$ dynamical strange quarks, the disconnected diagram depicts the effect of $s\bar{s}-gg(\cdots)$ transition during the propagation of the scalar meson. On the hadron level, this transition can be viewed as the glueball-$s\bar{s}$ meson mixing effect. The large mass change of $f_0$ hints at a very strong mixing effect in the scalar channel. So we include the glueball operator in the extraction of the scalar meson mass. We adopt the strategy in Refs.~\cite{Morningstar:1999rf,Chen:2005mg} to get the optimized glueball operator $\mathcal{O}_G(t)=\mathcal{O}^\dagger_G(t)$ coupling mainly to the ground state. Based on different prototypes of Wilson loops and different smearing schemes, we build an operator set $\mathcal{S}=\{\mathcal{O}_\alpha, \alpha=1,2,\ldots, 24\}$ with each operator $\mathcal{O}_\alpha$ belonging to the $A_1$ irreducible representation of the lattice symmetry group O, the octahedral group, and having the parity and the charge conjugate quantum numbers $\text{PC}=++$. We solve the generalized eigenvalue problem (GEVP) for this vacuum-subtracted matrix,
\begin{equation}\label{eq:gevp}
    C_{\alpha\beta}(t_d)w^{(i)}_\beta=\lambda^{(i)}(t_d-t_0) C_{\alpha\beta}(t_0)w^{(i)}_\beta,
\end{equation}
for a properly chosen $t_d-t_0$ (In practice, we set $t_0=0$ and $t_d=a_t$ to perform the GEVP analysis at this step), the eigenvector $w_\alpha$ of the largest eigenvalue $\lambda_\mathrm{max}(t_d-t_0)\sim e^{-m_\mathrm{min}(t_d-t_0)}$ gives the optimal glueball operator $\mathcal{O}^{(i)}_G=w^{(i)}_\alpha \mathcal{O}_\alpha$ that couples most to the lowest state of the quantum numbers $A_1^{++}$. In the continuum limit, $A_1$ subduces continuum spins $J=0,4,\ldots$. It is expected that the state of $J=4$ is higher than the $J=0$ state in mass, so we only consider the optimized gluonic operator $\mathcal{O}_G$ that couples most to the lowest state and treat it as a scalar operator in the continuum limit.

\begin{figure*}
\centering
\includegraphics[width=0.98\linewidth]{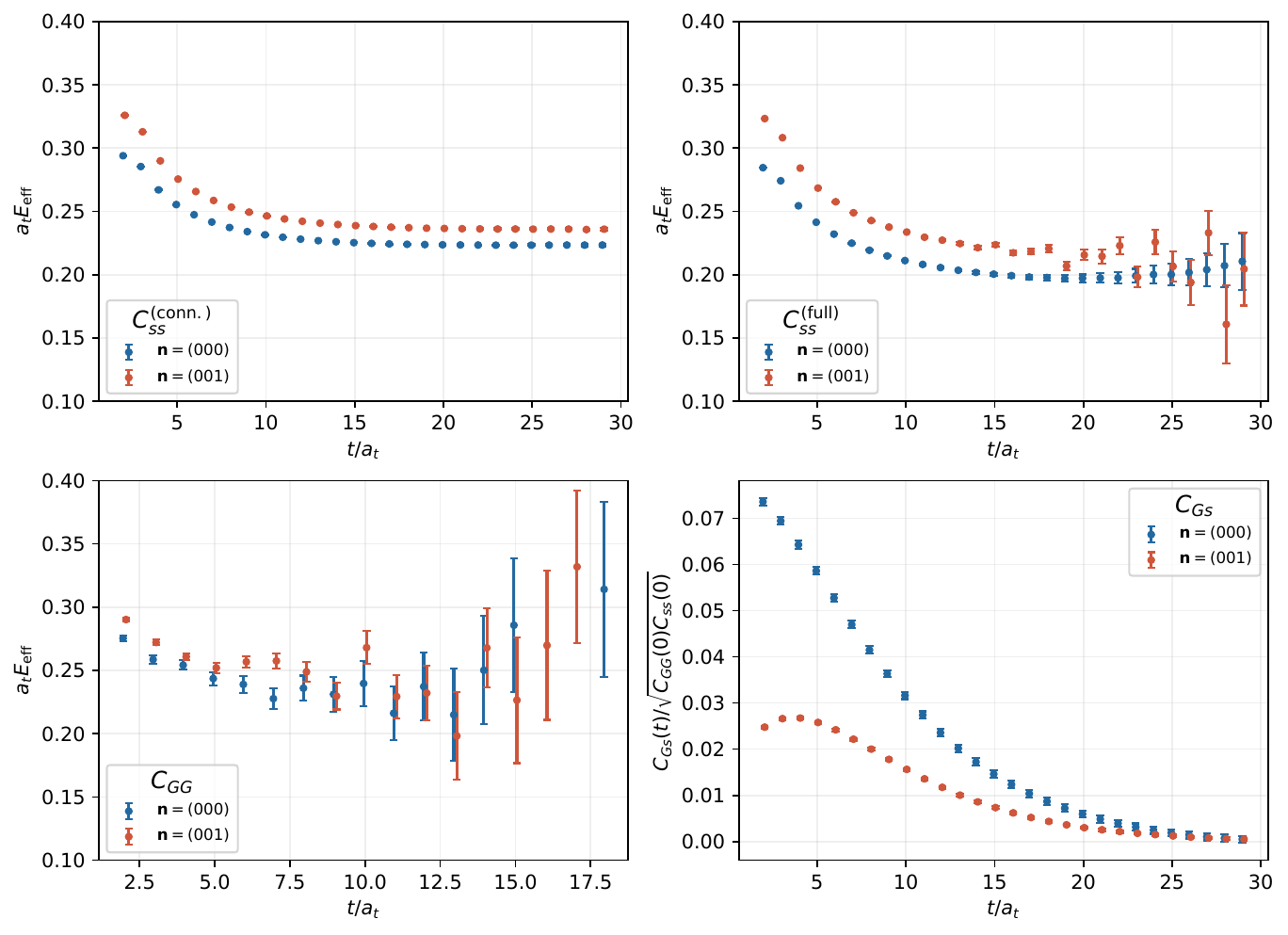}
\caption{Two-point functions at momentum modes $\vec n=(0,0,0)$ and $(0,0,1)$. The upper panels show effective energies from the connected-only scalar correlator $C_{ss}^{\rm (conn.)}(t)$ (left) and the full scalar correlator $C_{ss}^{\rm (full)}(t)$ (right); the lower-left panel is for the optimized gluonic correlator $C_{GG}$. Here $a_t E_\mathrm{eff}(t)=\ln \frac{C(t)}{C(t+a_t)}$. The lower right shows the correlator $C_{Gs}(t)$ normalized by $\sqrt{C_{GG}(0)C_{ss}(0)}$, with the operator phase chosen so that the displayed mixed correlator is positive. Errors are statistical.}
\label{fig:E-eff-001}
\end{figure*}

\section{  \texorpdfstring{$s\bar{s}$ and glueball mixing in $0^{++}$ channel}{s-sbar and glueball mixing in the scalar channel} }\label{sec:mixing1}
\subsection{Eigen energies}

We analyze the correlation matrix of the smeared scalar operator $\mathcal{O}_s=\bar\psi^{(s)}\psi^{(s)}$ and the optimized gluonic operator $\mathcal{O}_G$, namely,
\begin{equation}\label{eq:corr-sg}
 \mathbf C(t)=\begin{pmatrix}C_{GG}(t)&C_{Gs}(t)\\C_{sG}(t)&C_{ss}(t)\end{pmatrix},
\end{equation}
at momenta $\vec p=2\pi\vec n/(La_s)$ with $\vec n=(0,0,0)$ and $(0,0,1)$. At zero momentum both operators have nonzero vacuum expectation values (VEV). We use conventional vacuum subtraction,
\begin{equation}\label{eq:vacuumsub}
 C_{XY}(t)=\langle\mathcal{O}_X(t)\mathcal{O}_Y^\dagger(0)\rangle
 -\langle\mathcal{O}_X\rangle\langle\mathcal{O}_Y^\dagger\rangle,
 \qquad X,Y=G,s,
\end{equation}
where $\langle \mathcal{O}_X\rangle$ stands for the global VEV of $\mathcal{O}_X$ (Note that the VEV of $\mathcal{O}_G$ has been subtracted when it is derived in the first-step GEVP analysis described above). At nonzero momentum the vacuum expectation values vanish and the vacuum subtraction is not necessary.
Figure~\ref{fig:E-eff-001} compares the effective energies $a_t E_\mathrm{eff}(t)=\ln \frac{C(t)}{C(t+a_t)}$ at two momentum modes. The upper two panels are the effective energies of $C_{ss}(t)$ from the connected diagram contribution $C_{ss}^\mathrm{(conn.)}(t)$(left panel) and that of the full $C_{ss}^\mathrm{(full)}(t)$ with inclusion of the disconnected diagrams (right panel), respectively. It is seen that the inclusion of disconnected diagrams lowers the effective energies substantially. The lower left panel shows the effective energies of $C_{GG}$, which illustrate the goodness of the signal-to-noise ratios of the glueball operators. The lower right panel indicates the time dependence of $C_{Gs}(t)$.

\begin{figure}[htbp]
\centering
\includegraphics[width=0.95\linewidth]{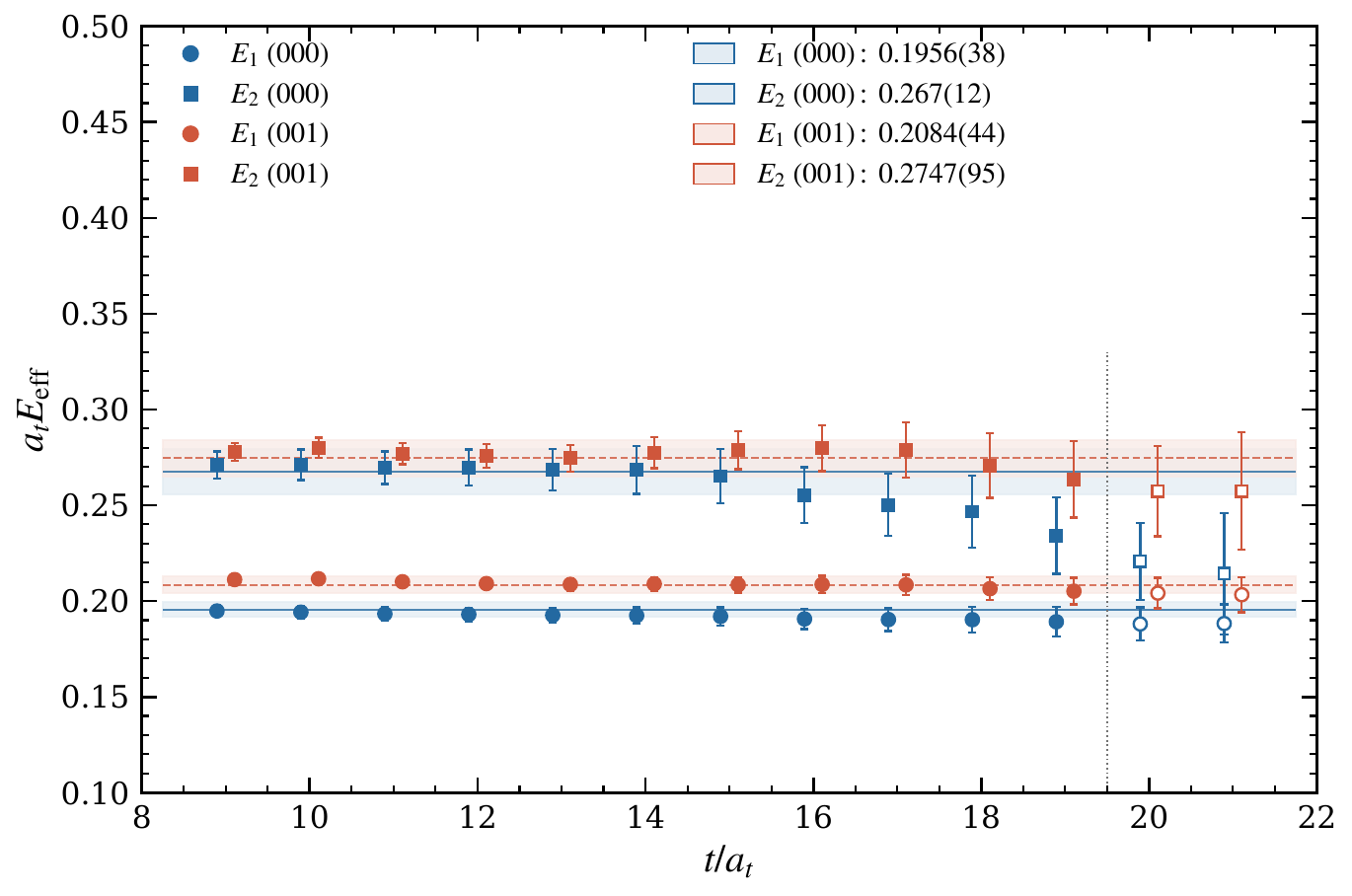}
\caption{Effective energies at momentum modes $(0,0,0)$ (blue) and $(0,0,1)$ (red). Points illustrate the effective energies $E_i^{\rm (eff)}$ defined by Eq.~(\ref{eq:eff-ene}) for the $(t_0,t_d)/a_t=(8,9)$ case. Lines and bands show the AIC estimates over $t_0/a_t=5$--$9$ and accepted fit windows, with the two uncertainty components added in quadrature.}
\label{fig:effective_lambda}
\end{figure}

We focus on the lowest two energy eigenstates (labelled by $|1\rangle$ and $|2\rangle$). In order to obtain their energies ($E_1$ and $E_2$), we perform the second-step GEVP analysis of $\mathbf C(t)$ according to Eq.~(\ref{eq:gevp}). For each choice of $t_0$, we use $t_d=t_0+a_t$ to determine the eigenvectors $(w_G^{(i)},w_s^{(i)})$ for $i=1,2$, which give the optimized operators
\begin{equation}\label{eq:opt-oper}
 \mathcal O^{(i)}=w_G^{(i)}\mathcal O_G+w_s^{(i)}\mathcal O_s,
 \qquad i=1,2.
\end{equation}
The correlator $\langle \mathcal{O}^{(i)}(t)\mathcal{O}^{(i),\dagger}(0)\rangle$, normalized at $t_0$, defines $\lambda_i(t,t_0)$, from which the effective energy is defined as
\begin{equation}\label{eq:eff-ene}
 E_i^{\rm eff}(t)=-\frac{1}{t-t_0}\ln\lambda_i(t,t_0).
\end{equation}
The data points in Fig.~\ref{fig:effective_lambda} illustrate the effective energies of the two momentum modes for $(t_0,t_d)/a_t=(8,9)$, where the error bars are from a bootstrap analysis. It is seen that fairly good plateaus show up when $t\ge 9$, so we extract the energies from single-exponential fits to $\lambda_i(t,t_0)$ over several time windows. In practice, we vary $t_0$ in the range $t_0/a_t=5$--$9$ and the fit intervals, and combine the accepted fits using AIC-type weights~\cite{Jay:2020jkz}. The uncertainty of the fitted energy is obtained through $\sigma=\sqrt{\sigma_{\rm stat}^2+\sigma_{\rm var}^2}$, where $\sigma_{\rm stat}$ and $\sigma_{\rm var}$ are the bootstrap statistical error of the AIC-averaged result and the AIC-weighted fit-window/$t_0$ variation, respectively. The colored bands in Fig.~\ref{fig:effective_lambda} illustrate the results of energies $E_{1,2}$ at the two momentum modes, whose values in physical units (converted from $a_t^{-1}=6.66$ GeV) are listed in Table~\ref{tab:mixing_results}, where $m_i(0,0,1)$ denotes the masses obtained from $E_i(0,0,1)$ using $m_i=\sqrt{E_i^2-\vec p^{\,2}}$. Obviously, the values of $m_i(0,0,1)$ agree with $E_i(0,0,0)$ (masses actually) very well. We combine these values by a correlated least-squares fit to two common masses, retaining correlations between momentum sectors and between the two states:
\begin{equation}\label{eq:masses}
 m_1=1.296(20)~\mathrm{GeV},\qquad
 m_2=1.768(48)~\mathrm{GeV}.
\end{equation}

\begin{table}[htbp]
\centering
\setlength{\belowcaptionskip}{2pt}
\caption{Energies and mixing angles in the rest and moving frame. The energies are in physical units (GeV) converted from $a_t^{-1}=6.66$ GeV, and the statistical uncertainty and fit-window/$t_0$ variation are combined in quadrature. The masses $m_i(0,0,1)$ are derived from $E_i(0,0,1)$ through the dispersion relation. The bottom row shows the final results combined from those in the two frames.}
\label{tab:mixing_results}
\begin{tabular}{cccc}
\toprule

    & $|1\rangle$ & $|2\rangle$ & $|\theta|$\\
\midrule

$E_i(0,0,1)$ & $1.388(30)$ & $1.829(63)$ & $41.8(2.5)^\circ$\\
$m_i(0,0,1)$ & $1.286(32)$ & $1.753(66)$ &              \\

$E_i(0,0,0)$ & $1.303(25)$ & $1.781(79)$ & $38.3(3.4)^\circ$\\
\midrule
$m_i$        & $1.296(20)$ & $1.768(48)$ & $40.5(2.2)^\circ$\\
\bottomrule

\end{tabular}
\end{table}

\subsection{Mixing angle}
\begin{figure}[htbp]
\centering
\includegraphics[width=0.95\linewidth]{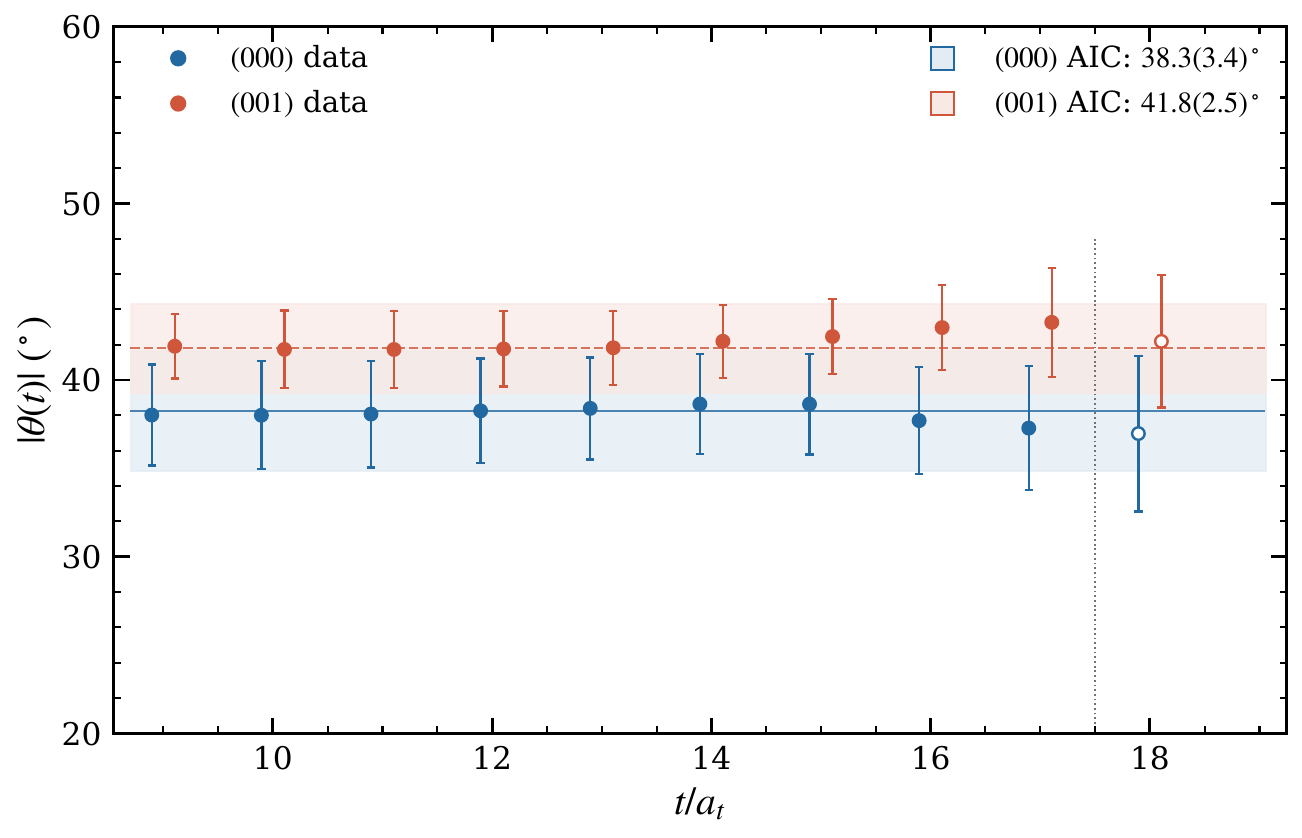}
\caption{Mixing angles at momentum modes $(0,0,0)$ and $(0,0,1)$. Points illustrate the mixing angle $|\theta(t)|$ defined in Eq.~(\ref{eq:angle}) for the $(t_0,t_d)/a_t=(8,9)$ case, with the errorbars being statistical. Lines and bands show the separate AIC estimates over the accepted fit windows and $t_0/a_t=5$--$9$, where the statistical uncertainty and fit-window/$t_0$ variation are combined in quadrature.}
\label{fig:theta_t}
\end{figure}

For each of the momentum modes of $\vec n=(0,0,0)$ and $(0,0,1)$, the matrix elements of the correlation matrix $\mathbf{C}(t)$ in Eq.~(\ref{eq:corr-sg}) have the following spectral expression
\begin{equation}\label{eq:spectral}
    C_{XY}(t)=\sum\limits_{i=1}^2 Z_X^i Z_Y^i\left(e^{-E_it}+e^{-E_i(T-t)}\right),
\end{equation}
where $Z_X^{i}=\langle 0|\mathcal{O}_X|i\rangle$. Let $|G\rangle$ and $|s\bar{s}\rangle$ be the pure scalar glueball and the pure $s\bar{s}$ states, respectively, the eigen states $|1\rangle$ and $|2\rangle$ can be viewed as their admixtures through a mixing angle $\theta$, namely,
\begin{equation}\label{eq:mixing}
    \left( \begin{array}{c} |1\rangle\\ |2\rangle \end{array} \right)
    =\left( \begin{array}{cc} \cos \theta & -\sin\theta \\
                              \sin \theta &  \cos \theta
            \end{array}\right)
    \left( \begin{array}{c} |s\bar{s}\rangle\\ |G\rangle \end{array}\right).
\end{equation}
Here $|s\bar{s}\rangle$ and $|G\rangle$ denote the unmixed basis states used in both this section and Sec.~3.3, while the full correlators, including disconnected contractions, describe the mixed energy eigenstates $|1\rangle$ and $|2\rangle$.
The smeared operator $\mathcal{O}_s$ is expected to couple most to the lowest $s\bar{s}$ state $|s\bar{s}\rangle$ and the optimized gluonic operator $\mathcal{O}_G$ couples most to the lowest glueball state $|G\rangle$, then with the assumption $\langle 0|\mathcal{O}_X|Y\rangle\approx  Z_X\delta_{XY}$ we have~\cite{Christ:2010dd,Dudek:2013yja}
\begin{equation}\label{eq:mte}
    \begin{array}{ll}
    Z_s^1=~Z_s \cos\theta & Z_s^2=Z_s \sin\theta,\\
    Z_G^1=-Z_G \sin\theta & Z_G^2=Z_G \cos\theta.
    \end{array}
\end{equation}
where $Z_s=\langle 0|\mathcal{O}_s|s\bar{s}\rangle$ and $Z_G=\langle 0|\mathcal{O}_G|G\rangle$. On the other hand, since the optimized operator $\mathcal{O}^{(i)}$ in Eq.~(\ref{eq:opt-oper}) couples predominantly to the $i$-th energy eigenstate $|i\rangle$, we have the following expressions for the correlation functions $C_{Xi}(t)$ (after the vacuum subtraction)
\begin{eqnarray}
    C_{Xi}(t)&\equiv&\langle \mathcal{O}_X(t)\mathcal{O}^{(i)\dagger}(0)\rangle \approx Z_X^i Z_i e^{-E_i t},
\end{eqnarray}
where $Z_i=\langle 0|\mathcal{O}^{(i)}|i\rangle$. Consequently, the mixing angle can be determined through the relation
\begin{equation}\label{eq:angle}
    |\tan\theta(t)|=\sqrt{-\frac{Z_s^2 Z_G^1}{Z_s^1 Z_G^2}}\approx \sqrt{-\frac{C_{s2}(t)C_{G1}(t)}{C_{s1}(t)C_{G2}(t)}}.
\end{equation}
The data points in Fig.~\ref{fig:theta_t} illustrate the obtained $|\theta(t)|$ in the range $t/a_t\in [9,18]$ for the case of $(t_0,t_d)/a_t=(8,9)$, where one can see that the values of $|\theta(t)|$ at the two momentum modes are insensitive to $t$ in this time range. Following the same AIC strategy as in Sec. 3.1 for the energies $E_1$ and $E_2$ (considering the variation of $t_0/a_t=5$--$9$ and the change of fit ranges), we obtain $|\theta|=38.3(3.4)^\circ$ for $\vec n=(0,0,0)$ and $|\theta|=41.8(2.5)^\circ$ for $\vec n=(0,0,1)$, respectively, which are consistent with each other. These values are also illustrated by colored bands in Fig.~\ref{fig:theta_t}. Combining these two values we determine the mixing angle to be
\begin{equation}\label{eq:common_angle}
 |\theta|=40.5(2.2)^{\circ}.
\end{equation}
The two momentum-sector results are combined by a correlated least-squares fit to a common angle. For both masses and the mixing angle, the quoted uncertainties combine the propagated statistical uncertainty and fit-window/$t_0$ variation in quadrature.

\subsection{A phenomenological consistency check}
As a phenomenological consistency check, we determine the mixing angle in the two-state model depicted in Eq.~(\ref{eq:mixing}) using the lattice results of masses. In the two-state space spanned by the pure $s\bar{s}$ state $|s\bar{s}\rangle$ and the pure glueball state $|G\rangle$, the Hamiltonian $H$ can be expressed as
\begin{equation}
    H=\left(\begin{array}{cc} m_{s\bar{s}} & x_s\\ x_s & m_G\end{array}\right),
\end{equation}
where $m_{s\bar{s}}$ and $m_G$ are the masses of $|s\bar{s}\rangle$ and $|G\rangle$ respectively, and $x_s$ is the mixing energy. For $\Delta=m_G-m_{s\bar{s}}$   and $\bar{m}=(m_{s\bar{s}}+m_G)/2$, the mass eigenvalues and the magnitude of the mixing angle  can be expressed as
  \begin{eqnarray}\label{eq:eigenmass}
 m_\pm&=&\bar m\pm\frac{1}{2}\sqrt{\Delta^2+4x_s^2},\nonumber\\
 |\sin\theta|&=&\sqrt{\frac{1}{2}\left(1-\frac{\Delta}{\sqrt{\Delta^2+4x_s^2}}\right)},
\end{eqnarray}
where $m_{1,2}=m_{\mp}$. Given the values of $m_{1,2}$ in Eq.~(\ref{eq:masses}), we need the value of $\Delta$ to determine $x_s$ and $|\sin\theta|$. According to the relation $m_1+m_2=m_G+m_{s\bar{s}}$, $\Delta$ is determined by $m_{s\bar{s}}$ through $\Delta=(m_1+m_2)-2m_{s\bar{s}}$. While the overlap analysis in Sec.~3.2 does not require the unmixed masses as inputs, the present mass-matrix analysis requires an estimate of $m_{s\bar{s}}$. Phenomenologically, a pure $s\bar{s}$ state is defined in the quark model and propagates without the annihilation of the constituent $s\bar{s}$ pair. So we follow the prescription used in earlier lattice studies~\mbox{
\cite{McNeile:2000xx}} to estimate the mass $m_{s\bar{s}}$ of the pure $s\bar{s}$ scalar meson by the mass obtained from the connected diagram (see in Table~\ref{tab:spectrum}), namely
\begin{equation}
 m_{s\bar{s}}\simeq m_{s\bar{s}}^{(C)}=1.486(2)~\mathrm{GeV},
\end{equation}
\hskip0pt
and obtain subsequently the mass difference $\Delta=92(64)~\mathrm{MeV}$.
Finally, we obtain the mixing energy and the mixing angle
\begin{equation}
|x_s|=231(14)~\mathrm{MeV},\quad |\theta|=39.4(3.6)^{\circ}
\end{equation}
from Eq.~(\ref{eq:eigenmass}), as well as the estimate of the pure scalar glueball mass
 \begin{equation}\label{eq:mg}
    m_G\approx 1.578(64)~\mathrm{GeV}.
\end{equation}
The resulting angle is in good agreement with the direct determination $|\theta|=40.5(2.2)^{\circ}$ and provides a phenomenological consistency check within this two-state prescription. The estimated $m_G$ is also close to the quenched lattice QCD result $m_G\approx 1.56\,\rm{GeV}$ at $a_s\approx 0.156\,\rm{fm}$~\cite{Chen:2005mg}.

\section{Discussion}\label{sec:discussion}
As for the phenomenological implication of the large mixing angle and the large mixing energy, we must consider the physical case that there are three light quark flavors ($u,d,s$ quarks) with the current quark masses satisfying $m_u\approx m_d=m_l\ll m_s$. Intuitively, the dynamics of this kind of mixing is necessarily dictated by the gluon-$q\bar{q}$ coupling, so we consider first the flavor number ($N_f$) dependence of the mixing in the SU($N_f$) symmetric limit and then discuss the SU($N_f$) symmetry-breaking effect owing to $m_l\ll m_s$.

In the QCD with $N_f=3$ degenerate dynamical light quarks of masses $m_u=m_d=m_s\equiv m_q$, the lowest scalar glueball $G$ mixes with the lightest flavor singlet scalar meson $f_0^1$ of the flavor wave function
\begin{equation}
    |f_0^1\rangle=\frac{1}{\sqrt{3}}(|u\bar{u}\rangle+|d\bar{d}\rangle+|s\bar{s}\rangle).
\end{equation}
In the limit of the SU(3) flavor symmetry, gluons couple equally to $u\bar{u}$, $d\bar{d}$ and $s\bar{s}$. When $m_q$ is set to be the physical strange quark mass $m_s$, the mixing energy $x(N_f=3)$ is expected to be $\sqrt{3}x_s(N_f=1)$ at the same quark mass $m_q=m_s$. For $x_s(N_f=1)=231(14)\,\rm{MeV}$ obtained in this work, we have $x_s(N_f=3)=401(24)\,\rm{MeV}$ for the $f_0^1-G$ mixing. There is actually a lattice QCD study with $N_f=3$ degenerate light quarks with $m_q$ being set to give $m_\pi=m_{\eta^8}= 688(1)\,\rm{MeV}$ at the lattice spacing $a_s\approx 0.148\,\rm{fm}$~\cite{Woss:2020ayi} that is almost the same as the $a_s$ in the present work. In this Reference, the mass of the flavor octet scalar meson $f_0^8$ is estimated to be $m_{f_0^8}\approx 2m_{\eta^8}\approx 1.38\,\rm{GeV}$. If this mass value is taken for the mass of the unmixed $f_0^1$ since it is derived from the connected diagram only, Eq.~(\ref{eq:eigenmass}) along with $m_G$ in Eq.~(\ref{eq:mg}) gives the masses of the two mass eigen states after the $f_0^1-G$ mixing as
\begin{equation}
   m_1=1.066(19)~\mathrm{GeV},\quad m_2=1.892(58)~\mathrm{GeV}.
\end{equation}
These are illustrative model estimates with the $1.38~\mathrm{GeV}$ input held fixed; its uncertainty and the change in sea-quark content are not included. It is interesting to see that $m_1$ is close to the measured mass value $0.934(8)\,\mathrm{GeV}$ in Ref.~\cite{Woss:2020ayi}, where this mass eigenstate is named $f_0^1$. This justifies, to some extent, that the $f_0^1-G$ mixing and the mixing energy $x_s$ we obtain are reasonable.

The improved gauge and fermion actions used in this work are expected to reduce discretization effects. Nevertheless, calculations at several lattice spacings are necessary to assess these effects and determine  the  continuum limit of   the mixing parameters. Such calculations require  large gauge ensembles on finer lattices   and remain challenging with our current computational  resources.

In the physical case that $m_l\ll m_s$ breaks the SU(3) flavor symmetry explicitly, the symmetry-breaking effect should be considered for the scalar $q\bar{q}-G$ mixing. So the quark mass dependence of the mixing energy $x_q$ (Here the subscript $q$ refers to the quark flavors) should also be explored. Reference~\cite{McNeile:2000xx} extracts $x_q$ at two quark masses $m_q\approx m_s$ and $m_q\approx 2m_s$ and does not observe a sizable difference in $x_q$. Reference~\cite{Lee:1999kv} observes the ratio $x_l/x_s\approx 1.2$ ($l$ referring to $u,d$ quarks) which is almost independent of the lattice spacing .

Theoretically, Chanowitz proposes that~\cite{Chanowitz:2005du,Chanowitz:2007ma}, to all orders in perturbative QCD, the amplitude $\mathcal{T}(G\to q\bar{q})$ is proportional to the {\it current} quark mass $m_q$ in the (pseudo)scalar channel. This is the so-called `chiral suppression'. Following the logic in Ref.~\cite{Chanowitz:2005du,Chao:2007sk}, the transition amplitude for $m_q\to 0$ in the perturbative QCD is expressed as
\begin{eqnarray}
    &&\mathcal{T}(G\to q(p_1)\bar{q}(p_2))\nonumber\\
    &&=\bar{u}_s(p_1)\left((A\slashed{p}_1+B\slashed{p}_2)+\mathcal{O}(m_q)\right)v_s(p_2),
\end{eqnarray}
where $u$ and $v$ are the spinors of the quark and antiquark, respectively, and $A$ and $B$ are two form factors. When the final-state quark and antiquark are free particles, namely, $(\slashed{p}-m_q)u_s(p)=0$ and $(\slashed{p}+m_q)v_s(p)=0$, one has $\mathcal{T}\propto m_q$, which vanishes in the chiral limit $m_q=0$. However, for the quark and antiquark that are bound in a scalar $q\bar{q}$ meson, they interact with each other and are necessarily off the mass shell. In the non-relativistic quark model picture, the valence quark in a hadron are dressed and have an effective mass, namely, the constituent mass $\tilde{m}_q$ due to the nonperturbative feature of QCD. As a qualitative illustration, if we take the approximation $(\slashed{p}-\tilde{m}_q)u_s(p)\approx 0$ and $(\slashed{p}+\tilde{m}_q)v_s(p)\approx 0$ for the quark spinors, we have
\begin{equation}
    \mathcal{T}(G\to q(p_1) \bar{q}(p_2))\approx (A-B)\tilde{m}_q+\mathcal{O}(m_q).
\end{equation}
Phenomenologically, one usually takes $\tilde{m}_{u,d}\sim 300$ MeV and $\tilde{m}_s\sim 400-500$ MeV, which imply a milder SU(3) symmetry breaking effect for the $G$-$q\bar{q}$ meson mixing.

Based on the arguments above, it is likely that the scalar glueball mixes strongly with $q\bar{q}$ scalar mesons, and the mixing strengths, measured by the mixing energy $x_q$, are comparable among $u\bar{u}$, $d\bar{d}$ and $s\bar{s}$. The ratio $\rho\equiv x_l/x_s$ is likely $\mathcal{O}(1)$, and its deviation from one measures the flavor SU(3) symmetry-breaking effect for the $G-q\bar{q}$ mixing. If the flavor wave function of a scalar $q\bar{q}$ meson is written as
\begin{equation}
    |f_0^{(\alpha)}\rangle=\frac{1}{\sqrt{2}}\left(|u\bar{u}\rangle+|d\bar{d}\rangle\right)\cos\alpha-|s\bar{s}\rangle\sin\alpha,
\end{equation}
then the $f_0^{(\alpha)}-G$ mixing energy is
\begin{equation}
    x^{(\alpha)}=\left(\rho\sqrt{2}\cos\alpha-\sin\alpha\right)x_s.
\end{equation}
The large value $x_s\sim 200$ MeV and $\rho\sim \mathcal{O}(1)$ have two-fold consequences: First, the lowest scalar glueball $G$ can mix with light $q\bar{q}$ scalar mesons in a wide mass range. Secondly, the physical mass spectrum of $f_0$ mesons can be very different from the spectrum of the pure $q\bar{q}$ mesons and the pure glueball $G$ before the mixing. Specifically speaking, if $f_0^{(\alpha)}$ is mainly a flavor octet, then $\alpha\sim 54.7^\circ$ ($\sin\alpha\sim \sqrt{2/3}$) results in a small mixing energy $x^{(\alpha)}$ and thereby a weak mixing with $G$. A mainly flavor singlet $f_0^{(\alpha)}$ has $\alpha\sim -35.3^\circ$ and mixes strongly with $G$.

The radiative $J/\psi$ decay is gluon-rich and is usually thought of as an ideal hunting ground for glueballs. According to the naive $\alpha_s$ power counting ($\alpha_s$ is the strong coupling constant here), the production of a $q\bar{q}$ meson in this process is $\mathcal{O}(\alpha_s^2)$ suppressed in comparison with that of a glueball. So the production rate of a scalar meson $f_0$ in the radiative $J/\psi$ decay can be used as a probe of its glueball component. In doing so, the latest lattice QCD calculation prediction (in the quenched approximation) $\mathrm{Br}(J/\psi\to \gamma G)$ to be $6.2(9)\times 10^{-3}$~\cite{Zou:2024ksc} (updated from the previous value $3.8(9)\times 10^{-3}$~\cite{Gui:2012gx}) can be used as a theoretical input. We first consider the possible glueball content of $f_0(1370)$, $f_0(1500)$ and $f_0(1710)$.

 $\mathbf{f_0(1710)}$ is observed in the $J/\psi$ radiative decays $J/\psi\to \gamma f_0(1710)\to \gamma(\pi\pi, K\bar{K},\eta\eta,\omega\omega,\omega\phi)$ and the production fractions are summed up to be larger than $2.1\times 10^{-3}$~\cite{ParticleDataGroup:2024cfk}, such that the fraction of its glueball content is expected to be larger than $30\%$ if its production is mainly from its glueball component. This large glueball content suggests that $f_0(1710)$ is mainly a flavor singlet. This is supported by its absence in the $\eta\eta'$ system that must be a flavor singlet. The partial wave analysis of the process $J/\psi\to \gamma\eta\eta'$ by the BESIII Collaboration gives an upper limit $\Gamma(\eta\eta')/\Gamma(\pi\pi)<2.87\times 10^{-3}$~\cite{BESIII:2022iwi}. However, BESIII also observes $f_0(1710)$ in flavor octet system $\omega\phi$ in the process $J/\psi\to \gamma \omega\phi$ and gives a fairly large branching fraction   $\mathrm{Br}(J/\psi\to\gamma f_0(1710)\to \gamma \omega\phi)=(2.5\pm 0.6)\times 10^{-4}$ ~\cite{BES:2006ssn, BESIII:2012rtd,ParticleDataGroup:2024cfk}. This needs to be understood.

$\mathbf{f_0(1500)}$ is observed in the $(\pi\pi, K\bar{K},\eta\eta)$ systems in the $J/\psi$ radiative decays with the summed production fraction $\mathrm{Br}(J/\psi\to\gamma f_0(1500))\approx 2.0\times 10^{-4}$~\cite{ParticleDataGroup:2024cfk}, which is almost an order of magnitude smaller than that of $f_0(1710)$. This is also the case for the individual combined branching fraction of $J/\psi\to\gamma f_0(1500/1710)\to\gamma PP$ obtained by BES and BESIII, where $PP$ refers to each of the $(\pi\pi, K_S K_S,\eta\eta)$ systems where both $f_0(1710)$ and $f_0(1500)$ are present~\cite{BES:2006ssn,BESIII:2018ubj,BESIII:2013qqz}. This indicates a tiny glueball content, implying that it is mainly a flavor octet. Furthermore, BESIII observes $f_0(1500)$ in $J/\psi\to\gamma\eta\eta'$ and the statistical significance is larger than $30\sigma$ and provides strong evidence that $f_0(1500)$ is not a flavor singlet. If mainly a $q\bar{q}$ meson, $f_0(1500)$ can be neatly filled in the scalar octet made up of $(a_0(1450), K_0^*(1430), f_0(1500))$. The SU(3) flavor symmetry requires the squared coupling constants for $f_0(1500)$ decaying into $(\pi\pi, K\bar{K},\eta\eta,\eta\eta')$ have relation 3:1:1:4, which gives the decay fractions satisfying $$\Gamma_{\pi\pi}:\Gamma_{K\bar{K}}:\Gamma_{\eta\eta}:\Gamma_{\eta\eta'}=1:0.258_{-4}^{+2}:0.236_{-4}^{+4}:0.29_{-10}^{+8}$$ using the resonance parameters   $(m,\Gamma)=(1522(25),108(33))\,\rm{MeV}$  of $f_0(1500)$, which is in good agreement with the experimental result $$\Gamma_{\pi\pi}:\Gamma_{K\bar{K}}:\Gamma_{\eta\eta}:\Gamma_{\eta\eta'}=1:0.25(4):0.17(4):0.17(4)$$ from the experimental branching fractions $34.5(2.2)\%$, $8.5(1.0)\%$ and $6.0(9)\%$ for $\pi\pi,K\bar{K},\eta\eta$ decay modes~\cite{ParticleDataGroup:2024cfk}, respectively, as well as $\mathrm{Br}(\eta\eta')/\mathrm{Br}(\pi\pi)=0.166_{-40}^{+42}$ measured by BESIII~\cite{BESIII:2022iwi}.

$\mathbf{f_0(1370)}$ is less established than $f_0(1710)$ and $f_0(1500)$ (see reviews in PDG 2024~\cite{ParticleDataGroup:2024cfk} and Ref.\cite{Pelaez:2025wma}). Its existence was recently confirmed by a dispersive analysis of $\pi\pi$ scattering data~\cite{Pelaez:2022qby}. $f_0(1370)$ is observed in the $J/\psi\to\gamma(\pi\pi,K\bar{K})$ processes. PDG 2024 lists the branching fraction $\mathrm{Br}(J/\psi\to\gamma f_0(1370)\to\gamma K\bar{K})=4.2(1.5)\times 10^{-4}$~\cite{ParticleDataGroup:2024cfk,Dobbs:2015dwa}, but BESIII measures $\mathrm{Br}(J/\psi\to\gamma f_0(1370)\to\gamma K_SK_S)=1.1(4)\times 10^{-5}$~\cite{BESIII:2018ubj}, which is almost ten times smaller. The JPAC Collaboration reanalyses the $J/\psi\to\gamma (\pi\pi,K_SK_S)$ data by BESIII and identifies no $f_0(1370)$ but the four scalars $f_0(1500), f_0(1710), f_0(2020)$ and $f_0(2330)$ with the production rate of $f_0(1710)$ being much larger than that of $f_0(1500)$~\cite{Rodas:2021tyb}. In contrast, Sarantsev {\it et al.} perform a coupled-channel analysis of BESIII data on $J/\psi\to\gamma(\pi\pi,K_SK_S,\eta\eta,\omega\phi)$~\cite{Sarantsev:2021ein} and observe $f_0(1370)$ with the branching ratios   $\mathrm{Br}(J/\psi\to\gamma f_0(1370)\to \gamma\pi\pi)=3.8(1.0)\times 10^{-4}$  and $\mathrm{Br}(J/\psi\to\gamma f_0(1370)\to \gamma K\bar{K})=1.3(4)\times 10^{-4}$ respectively. So the production rate of   $f_0(1370)$  in the $J/\psi$ radiative decay is still an open question.

Specifically, in the analysis in Ref.~\cite{Sarantsev:2021ein}, the data set includes the BESIII data on $ J/\psi\to\gamma(\pi\pi,K_SK_S,\eta\eta,\omega\phi)$, the CERN SPS data on $\pi\pi,\eta\eta,\eta\eta'$, the BNL data on $\pi\pi\to K_S K_S$, the LEAR $p\bar{p}$ annihilation data. The coupled channel analysis requires ten scalar mesons: the mostly flavor octet particles $f_0(980)$, $f_0(1500)$, $f_0(1770)$,   $f_0(2100)$ , $f_0(2330)$, and the mostly flavor singlet states $f_0(500)$, $f_0(1370)$, $f_0(1710)$, $f_0(2020)$, $f_0(2200)$. It is observed that the production rates of these scalar mesons in the $J/\psi$ radiative decay exhibit a bump around $1.8\,\mathrm{GeV}$ with respect to the masses of these states. By assuming a scalar glueball is distributed among these scalar mesons and through a Breit-Wigner amplitude to the distribution of the production rates, the glueball parameters are fitted to be $(m_G,\Gamma_G)=(1865\pm 25_{-30}^{+10}, 370\pm 50_{-20}^{+30})\,\rm{MeV}$, and the integrated production rate of these scalar mesons is determined to be $(5.8\pm 1.0)\times 10^{-3}$, which is amazingly consistent with the latest lattice QCD prediction of the glueball production rate $\mathrm{Br}(J/\psi\to\gamma G)=6.2(9)\times 10^{-3}$~\cite{Zou:2024ksc}. In a followup study, Klempt and Sarantsev assume the octet states are produced in $J/\psi$ radiative decays through their glueball components and the glueball contents of singlet states can be identified through the enhancement over a continuum background, and then determine the glueball fractions $\sin^2 \Phi_G$ in these states~\cite{Klempt:2021wpg,Gross:2022hyw}. It turns out that $f_0(1710)$,$f_0(1770)$,$f_0(2020)$ and $f_0(2100)$ have large glueball contents of $12\pm 6\%$,$25\pm10\%$,$16\pm 9\%$ and $17\pm 8\%$, respectively.
Although a lot of theoretical assumptions are mentioned in the data analysis above, the major feature is that the scalar glueball spreads over a wide mass region and is embodied as part of a lot of scalar mesons. This is consistent with the large $G-q\bar{q}$ mixing energy we obtain. The large glueball contents of the assumed octet states $f_0(1770)$ and $f_0(2100)$ are surprising and could be attributed to the SU(3) flavor symmetry breaking effect in the $G-q\bar{q}$ mixing. On the other hand, as we have addressed previously, the large mixing energy can alter the spectrum of the singlet scalar mesons, so it is questionable that they can be reasonably arranged on the Regge trajectory as was done in Ref.~\cite{Sarantsev:2021ein}.

\section{ Summary}\label{sec:summary}
We perform the first lattice QCD study on mixing of the scalar glueball $G$ and the $q\bar{q}$ meson with $N_f=1$ dynamical strange quarks.   Combining the zero- and nonzero-momentum GEVP ratio analyses gives a mixing angle  $|\theta|=40.5(2.2)^{\circ}$   and the eigenstate masses $m_1=1.296(20)\,\rm{GeV}$ and $m_2=1.768(48)\,\rm{GeV}$. Within the phenomenological two-state description of Sec.~3.3, where the diagonal $s\bar{s}$ mass parameter is identified with the connected-only scalar mass, we obtain a model-dependent mixing energy $|x_s|=231(14)\,\rm{MeV}$. The quoted uncertainties combine statistical uncertainty and fit-window/$t_0$ variation in quadrature. The calculation is performed at one lattice spacing and one spatial volume, so the continuum and infinite-volume limits remain to be quantified. In addition, our variational analysis is restricted to gluonic and $s\bar{s}$ interpolating operators. Contributions from other states may not be fully removed with this limited operator basis and can introduce systematic uncertainties.

If the flavor symmetry holds to some extent in the physical case that has three flavors of light quarks $u,d,s$, then the mixing energy will be even larger for the mixing between the scalar glueball and the flavor singlet $q\bar{q}$ scalar mesons. Therefore, the mixing can take place in a wide mass region. This seems in accordance with the recent coupled channel analyses on data of the $J/\psi$ radiative decays in Ref.~\cite{Sarantsev:2021ein,Klempt:2021wpg}, which show that the lowest scalar glueball is spread as part of a lot of scalar mesons in a wide mass region around $1.8\,\mathrm{GeV}$.

In order to clarify the mechanism of the $G-q\bar{q}$ mixing, the quark mass dependence of the mixing energy should be explored after a careful continuum limit extrapolation. Although numerically challenging, this is a crucial mission to be fulfilled in the future.

\section{Declaration of competing interest}
The authors declare that they have no known competing financial interests or personal relationships that could have appeared to influence the work reported in this paper.

\section{Acknowledgements}
    This work is supported by the National Key Research and Development Program of China under Grant No.\,2025YFB3003602
    and the National Natural Science Foundation of China (NNSFC) under Grant Nos. 12293060, 12293065, 12175063, 12192264, and 12675103. WS is also supported by Chinese Academy of Sciences under Grant No.\,YSBR-101. The Chroma software system\,\cite{Edwards:2004sx}, QUDA library\,\cite{Clark:2009wm,Babich:2011np},
    and PyQUDA package\,\cite{jiang2024usequdalatticeqcd} are acknowledged. The computations were performed on the HPC clusters at the Institute of High Energy Physics (Beijing), China Spallation Neutron Source (Dongguan), xiang-jiang HPC clusters at Hunan Normal University (Changsha), and the ORISE computing environment.

 \bibliographystyle{elsarticle-num}
  \bibliography{references}

\end{document}